\documentclass{cimento}

\usepackage{graphicx}  
\usepackage{caption}
\usepackage{subcaption}
\usepackage{newtxmath}
\usepackage{tikz}
\usepackage[compat=1.1.0]{tikz-feynman}
\usetikzlibrary{decorations.markings}
\usetikzlibrary{shapes.misc}

\title{On the study of inclusive semileptonic decays of $B_s$-meson from lattice QCD}
\author{P.~Gambino\from{ins:x}\ETC,
S.~Hashimoto\from{ins:y},
S.~M\"achler\from{ins:x}\from{ins:z},
M.~Panero\from{ins:x},
F.~Sanfilippo\from{ins:r},
S.~Simula\from{ins:r},
A.~Smecca\from{ins:x}
\atque
N.~Tantalo\from{ins:t}\thanks{Presented by A.~Smecca at ``Incontri di Fisica delle Alte Energie 2023'' in Catania}}
  \instlist{\inst{ins:x} Dipartimento di Fisica, Universit\`a di Torino \& INFN, Sezione di Torino - Torino, Italy
    \inst{ins:y}  Theory Center, Institute of Particle and Nuclear Studies, High Energy Accelerator Research Organization
    (KEK) - Tsukuba, Japan
    \inst{ins:z} Physikinstitut, Universit\"at Z\"urich - Z\"urich, Switzerland
    \inst{ins:r} INFN, Sezione di Roma Tre - Rome, Italy
  \inst{ins:t} Dipartimento di Fisica, Università di Roma ``Tor Vergata'' \& INFN, Sezione di Roma ``Tor Vergata'' - Rome, Italy}

\begin{document}

\maketitle

\begin{abstract}
  In this contribution we describe a recent study, published in ref.~\cite{Gambino:2022dvu}, focused on the lattice calculation of inclusive decay rates of heavy mesons. We show how the inclusive calculation can be achieved starting from four-point lattice correlation functions normalised appropriately. The correlators used in this project come from gauge ensembles provided by the JLQCD and ETM collaborations. An essential point of this method is the extraction of spectral densities from lattice correlators which is obtained using two of the most recent approaches in the literature. Our results are in remarkable agreement with analytical predictions from the operator-product expansion.
  This study represents the first step towards a full lattice QCD study of heavy mesons inclusive semileptonic decays.
\end{abstract}

\section{Introduction}

One of the most interesting objects in quark flavour physics is the Cabibbo-Kobayashi-Maskawa (CKM) matrix. It is a complex unitary matrix in the Standard Model (SM) where the modulus of each of its elements parametrises the weak decays of quarks. The elements of the CKM matrix are fundamental parameters of the SM and cannot be directly calculated from theory alone. In order to determine the value of these matrix elements, one needs to combine experimental measurements of certain observables involving the weak decay of quarks together with precise theoretical calculations of some quantity parametrising the decay. By determining each CKM matrix element it is possible to verify if the CKM matrix is indeed unitary as the SM predicts, knowing that any deviation from unitarity would be indirect evidence for beyond the Standard Model (BSM) physics.


At the moment, there is a persistent tension between the inclusive and exclusive determination of two CKM matrix elements, $i.e.$ $|V_{ub}|$ and $|V_{cb}|$. These determinations are obtained studying exclusive or inclusive semileptonic decays of $B$-mesons, where in the first case the $B$-meson decays to a specific daughter meson plus the leptonic pair while in the latter the $B$-meson decays to any possible final states allowed by the conservation of quantum numbers plus the leptonic pair.
In fact, this tension might not be a direct result of new physics, as it has been argued that BSM models struggle to accommodate this discrepancy in a consistent and significant way~\cite{Jung:2018lfu,Crivellin:2014zpa}.

In order to resolve this tension it is important to better understand the analysis behind the exclusive and inclusive determinations of these CKM matrix elements. From the theory side, the exclusive determination requires the calculation of non-perturbative form factors which nowadays can be calculated very precisely from lattice QCD simulations~\cite{FlavourLatticeAveragingGroup:2019iem}, while the computation of inclusive quantities used the operator product expansion (OPE) technique~\cite{WilsonOPE, KadanoffOPE}.

In this contribution we show a method that can be used to calculate inclusive quantities using lattice QCD correlation functions.

\section{Theoretical framework}
Following the mathematical formalism introduced in ref.~\cite{GambinoHashimoto}, we focus on the inclusive semileptonic decay rate of a $B_s$ meson decaying into some charmed final state $X_c$ and a pair of leptons $l\overline{\nu}$. Choosing the rest frame of the $B_s$ meson, one can write the differential decay rate as
\begin{equation}
  \frac{d\Gamma}{dq^2dq^0dE_l}=\frac{G_F^2|V_{cb}|^2}{8\pi^3}L_{\mu\nu}W^{\mu\nu},
\end{equation}
where $L_{\mu\nu}$ and $W_{\mu\nu}$ are respectively the leptonic and hadronic tensor. It is useful to write the hadronic tensor in its spectral representation as
\begin{equation}
  W_{\mu\nu}(\omega,q)=\frac{(2\pi)^3}{2M_{B_s}}\langle \overline{B}_s(\boldsymbol{0})|J^{\dagger}_{\mu}(0)\delta(\hat{H}-\omega)\delta^3(\hat{\boldsymbol{P}}-\boldsymbol{q})J_{\nu}(0)|\overline{B}_s(\boldsymbol{0})\rangle,  
\end{equation}
with the Hamiltonian operator $\hat{H}$ and momentum operator $\hat{\boldsymbol{P}}$ written explicitly.

After integrating analytically over $E_l$, the differential decay rate can be rewritten as
\begin{equation}
  \frac{d\Gamma}{d\boldsymbol{q}^2}=\frac{G_F^2|V_{cb}|^2}{24\pi^3|\boldsymbol{q}|}\sum_{l=0}^2\Big(\sqrt{\boldsymbol{q}^2}\Big)^{2-l}Z^{(l)}(\boldsymbol{q}^2),\hspace{.3cm}\text{with} \hspace{0.3cm}  Z^{(l)}(\boldsymbol{q}^2)=\int_{0}^{\infty}d\omega~\Theta^l(\omega_{\max}-\omega)Z^{(l)}(\omega,\boldsymbol{q}^2),
  \label{eq:Zinteg}  
\end{equation}
where $Z^{(l)}(\boldsymbol{q}^2)$ is the energy integral the hadronic tensor decomposed into Lorentz invariant structure functions $Z^{(l)}(\omega,\boldsymbol{q}^2)$.
The integration kernel $\Theta^l$ is defined as $\Theta^l(x)=x^l\theta(x)$, where $\theta(x)$ is the Heaviside step function, and it enforces the correct integration over the allowed phase space.

Eq.~\ref{eq:Zinteg} is the key quantity which allows to unlock the differential decay rate calculation. In the following section we will show how to compute it using lattice correlators.

\section{Lattice computation}
\begin{figure}
  \centering
  \begin{tikzpicture}
    [
      roundnode/.style={circle, draw=black, very thick, minimum size=1mm},
      squarednode/.style={rectangle, draw=blue!60, very thick, minimum size=6mm},
      cross/.style={cross out, draw=black, minimum size=4mm,very thick},
      decoration={
        markings,
    mark=at position 0.5 with {\arrow{stealth}}}
    ]
    \node[roundnode]      (Bmeson)                  {$B_s$};
    \node[cross, label={[yshift=-0.15cm,xshift=3mm]$J_{\mu}^{\dagger}$}] (current_d) [above right =.7cm and 1.7cm of Bmeson] {};
    \node[cross, label={[yshift=-0.05cm,xshift=-3mm]$J_{\nu}$}]  (current) [above right =.7cm and 6.2cm of Bmeson] {};      
    \node[roundnode] (Dmeson)       [right=8cm of Bmeson] {$B_s$};

    \draw[very thick,postaction={decorate}] (current_d.west) .. controls +(left:1.0cm) and +(up:3mm)  .. (Bmeson.north) node[midway,above]  {$b$};
    \draw[very thick,postaction={decorate}] (Bmeson.south) .. controls +(down:0.9cm) and +(down:0.9cm)  .. (Dmeson.south) node[midway,above] {$\bar{s}$};
    \draw[very thick,postaction={decorate}] (Dmeson.north) .. controls +(up:3mm) and +(right:1.0cm)  .. (current.east) node[midway,above] {$b$};
    \draw[very thick,postaction={decorate}] (current.west) -- (current_d.east) node[midway,above] {$c$};
    \draw [dashed] (2.25,0.5) -- (2.25,2.) node[left]{$t_2$};
    \draw [dashed] (6.75,0.5) -- (6.75,2.) node[right]{$t_1$};
    \draw [dashed] (0,-1.) -- (0,1.5) node[left]{$t_{\mathrm{snk}}$};
    \draw [dashed] (9,-1.) -- (9,1.5) node[right]{$t_{\mathrm{src}}$};

  \end{tikzpicture}
  \caption{Schematic representation of the four-point Euclidean correlation function defined in eq.~(\ref{eq:4pt}). The crosses represent the insertions of the weak currents at times $t_1$ and $t_2$, the meson states are created at time $t_{\mathrm{src}}$ and annihilated at time $t_{\mathrm{snk}}$. Between the currents we have the propagation of the charm quark, hence, the piece of the correlation functions defined between the currents contained all the possible charmed states $X_c$.}
  \label{fig:4pt}
\end{figure}
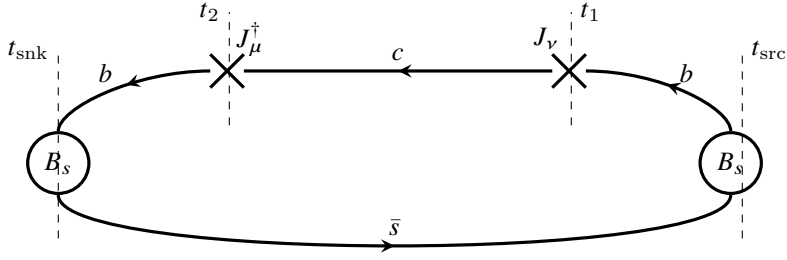

In ref.~\cite{GambinoHashimoto}, the authors show that in order to access the full spectrum of charmed final states, one needs to compute a four-point lattice correlation function, which can be written explicitly as
\begin{equation}
  C_{\mu\nu}(t_{\mathrm{snk}},t_2,t_1,t_{\mathrm{src}})=\int d^3x~e^{i\boldsymbol{q}\cdot \boldsymbol{x}}T\langle 0|\tilde{\phi}_{B_s}(\boldsymbol{0},t_{\mathrm{snk}})J^{\dagger}_{\mu}(\boldsymbol{x},t_2)J_{\nu}(\boldsymbol{0},t_1)\tilde{\phi}^{\dagger}_{B_s}(\boldsymbol{0},t_{\mathrm{src}})|0\rangle,
  \label{eq:4pt}
\end{equation}
where the two currents are sandwiched between the $B_s$ meson states, as shown in fig.~\ref{fig:4pt}.
The above equation can then be normalised with two-point correlators $C(t)$ in order to remove the contribution coming from the creation/annihilation of the $B_s$ meson,
\begin{equation}
  M_{\mu\nu}(t_2-t_1;\boldsymbol{q})=\lim_{\substack{t_{\mathrm{snk}}\to+\infty\\t_{\mathrm{src}}\to-\infty}}\frac{C_{\mu\nu}(t_{\mathrm{src}},t_2,t_1,t_{\mathrm{snk}})}{C(t_{\mathrm{snk}}-t_2)C(t_1-t_{\mathrm{src}})}.
  \label{eq:norm_corr}
\end{equation}

Then, it is possible to rewrite this expression as the Laplace transform of the hadronic tensor
\begin{equation}
  M_{\mu\nu}(t;\boldsymbol{q})=\int_0^{\infty}d\omega~W_{\mu\nu}(\omega,\boldsymbol{q}^2)e^{-\omega t},
\end{equation}
where here the time $t$ is understood as the time separation between the two currents $t=t_2-t_1$.

In our analysis we make use of the decomposition of the hadronic tensor in structure functions and we write a linear combination of normalised four-point lattice correlators
\begin{equation}
  G^{(l)}(a\tau;\boldsymbol{q})=\int_0^{\infty}Z^{(l)}_L(\omega,\boldsymbol{q}^2)~e^{-a\tau\omega},
  \label{eq:inv}
\end{equation}
where we used $t=a\tau$, $a$ being the lattice spacing.
The problem of computing eq.~\ref{eq:Zinteg} is then reduced to the problem of extracting $Z^{(l)}_L(\omega,\boldsymbol{q}^2)$ from eq.~\ref{eq:inv} and then performing the integral with the correct integration kernel.
This is known in the literature as an ill-posed inverse problem as the lattice correlators are unavoidably affected by statistical errors and are limited by the finite temporal size of the lattice.

In order to overcome the inverse problem, we employed two slightly different techniques which are known in the literature as the HLT method~\cite{HLT} and the Chebyshev polynomials method~\cite{Chebyshev}.

The correlators used in our work are obtained from two distinct ensembles of gauge configurations, one provided by the ETM collaboration~\cite{ETM1,ETM2} and the other one from the JLQCD collaboration~\cite{JLQCD1,JLQCD2}, using two different fermion discretisations namely Twisted Mass Wilson fermions and Domain-Wall fermions respectively. The valence quark in the ETM correlators were simulated using the Osterwalder-Seiler action~\cite{OS}.
For both ensembles, the $b$-quark mass (and hence the simulated $B_s$ meson) is unphysically light. This naturally affects the phase space region which can be accessed in the lattice calculation.

\section{Kernel reconstruction}
In our study we employed the Chebyshev polynomials method and HLT method on the JLQCD correlators and the ETMC correlators respectively. The first step in both methods is to convolute the spectral density (in our case $Z^{(l)}_L(\omega,\boldsymbol{q}^2)$) with a smooth kernel. This is done for two reasons: First, in order to overcome the ill-posed inverse problem it is necessary to reconstruct the integration kernel numerically, which can only work with sufficiently smooth functions. Second, the energy spectrum contained in a lattice correlator is a finite distribution of $\delta$-functions due to the finiteness of the volume in the simulation. Hence, in order to make contact with the physical quantity, one needs a continuous smooth function which can be extrapolated to infinite volume.

For this reason, in order to extract $Z^{(l)}_L(\omega,\boldsymbol{q}^2)$ one would normally choose a smeared version of a $\delta$-function. However, considering that our target quantity to access the differential decay rate is $Z^{(l)}(\boldsymbol{q}^2)$ (as shown in eq.~\ref{eq:Zinteg}), one can instead choose a smooth version of the integration kernel $\Theta^l_{\sigma}(\omega_{\max}-\omega)$, where $\sigma$ is a smearing parameter which will be removed at the end of the analysis. The kernel is then reconstructed in terms of a series of polynomials
\begin{equation}
  \Theta^l_{\sigma}(\omega_{\max}-\omega)=(\omega_{\max}-\omega)^l\theta_{\sigma}(\omega_{\max}-\omega)\simeq m_{B_s}^l\sum_{\tau}^{\tau_{\max}}g_{\tau}(\omega_{\max};\sigma)e^{-a\omega\tau},
\end{equation}
where $\tau_{\max}$ is the maximum time extent of the lattice correlator.

Once the coefficients $g_{\tau}$ have been obtained, either employing the Chebyshev polynomials or the HLT method (we encourage the interested reader to see refs.~\cite{HLT,Chebyshev} for details), we are able to apply them to the lattice correlator in order to compute $Z^{(l)}_{\sigma}(\boldsymbol{q}^2)$
\begin{align}
  Z^{(l)}_{\sigma,L}(\boldsymbol{q}^2)&=\int_0^{\infty}d\omega~\Theta^l_{\sigma}(\omega_{\max}-\omega)Z^{(l)}_L(\omega,\boldsymbol{q}^2)\\
  &\simeq \sum_{\tau}^{\tau_{\max}}g_{\tau}(\omega_{\max};\sigma)\int_0^{\infty}d\omega~Z^{(l)}_L(\omega,\boldsymbol{q}^2)e^{-a\omega\tau}
  \simeq \sum_{\tau}^{\tau_{\max}}g_{\tau}(\omega_{\max};\sigma)G^{(l)}(a\tau).
\end{align}

\section{Lattice results and comparison with the OPE}
As stressed in the previous section, in order to make contact with any physical quantity it is important to remove the dependence on the finite volume and the smearing. In particular, one needs to \textit{first} perform the infinite volume extrapolation and only then take the $\sigma \rightarrow 0$ limit:
\begin{equation}
  Z^{(l)}(\boldsymbol{q}^2)=\lim_{\sigma \rightarrow 0}\Bigg(\lim_{L\rightarrow \infty}Z^{(l)}_{\sigma,L}(\boldsymbol{q}^2)\Bigg).
\end{equation}
The two limits do not commute due to the fact that the infinite volume extrapolation is well-defined for continuous (smeared) quantities only. However, in our study we were unable to perform the infinite-volume extrapolation due to the fact that our data were obtained from simulations at only one physical volume. We quote our final results performing only the $\sigma \rightarrow 0$ limit, a choice which is justified considering that our present statistical uncertainties are likely to be larger than finite volume effects.

\begin{figure}
  \centering
  \includegraphics[width=.6\textwidth]{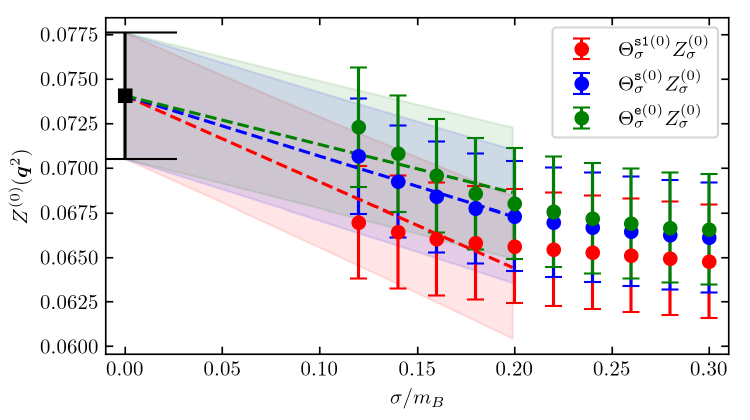}
  \caption{Combined $\sigma \rightarrow 0$ extrapolation of $Z^{(l)}_{\sigma}$ for the ETMC correlators, employing $10$ values of $\sigma \in [0.12m_{B_s},0.3m_{B_s}]$ and using the smallest $5$ to perform the fit}
  \label{fig:sig_zero}  
\end{figure}

An example of the $\sigma \rightarrow 0$ extrapolation is shown in fig.~\ref{fig:sig_zero}, where we show a combined linear fit of the results obtained with different versions of the smeared kernel $\Theta^{(l)}_{\sigma}$, for details see ref.~\cite{Gambino:2022dvu}.

\begin{figure}[t!]
  \centering
  \begin{subfigure}[t]{.5\textwidth}
    \centering
    \includegraphics[width=1.3\textwidth]{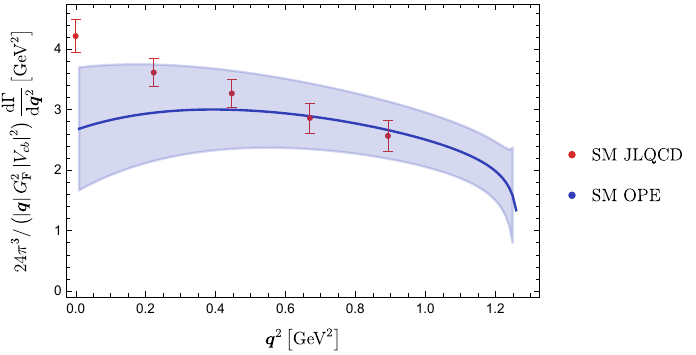}
  \end{subfigure}
  \begin{subfigure}[t]{.5\textwidth}
    \centering
    \includegraphics[width=1.3\textwidth]{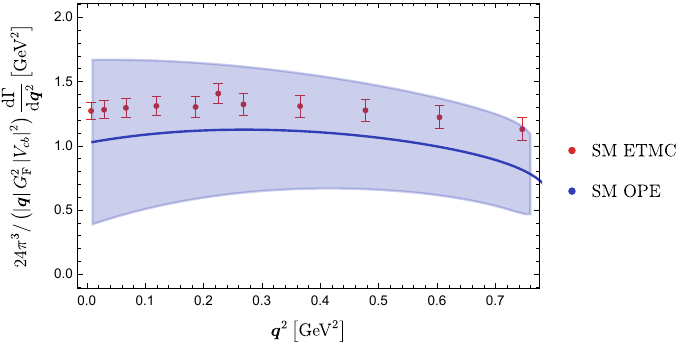}
  \end{subfigure}
  \caption{Differential $\boldsymbol{q}^2$ spectrum, divided by $|\boldsymbol{q}|$, in the SM. Comparison of OPE with JLQCD (top panel) and ETMC (bottom panel) data are shown.}
  \label{fig:OPE}  
\end{figure}

Finally, we are also able to compare the lattice results with the analytic predictions of the OPE. The two lattice results cannot be directly compared because they use different quark masses in their respective simulations. In fig.~\ref{fig:OPE} we see a remarkable agreement between the lattice results both in the JLQCD and the ETMC case. It is important to note that the uncertainty in the OPE is larger than the lattice one because of the unphysically light mass of the $b$-quark which enters the analysis through a $\frac{1}{m_b}$ expansion.

These results provide a non-trivial test for the method discussed in this work, making us optimistic that a full lattice QCD study including all the sources of systematic errors can be achieved in the near future. This is certainly a remarkable first step towards a better comprehension of the inclusive analysis with the hope that it could one day resolve the tension that affects the determination of $|V_{cb}|$.

\acknowledgments
A.~S. warmly thanks F. Giannuzzi and the conveners of the ``Intensity frontier'' session for the invitation to present this work and the organisers of the conference for the intellectually stimulating environment.


\begin{thebibliography}{0}
\bibitem{Gambino:2022dvu} \BY{Gambino P. \textit{et al.}} \IN{JHEP}{07}{2022}{083};
\bibitem{Jung:2018lfu} \BY{Jung M. \atque Straub D.~M.} \IN{JHEP}{01}{2019}{009};
\bibitem{Crivellin:2014zpa} \BY{Crivellin A. \atque Pokorski S.} \IN{Phys. Rev. Lett.}{114}{2015}{1};
\bibitem{FlavourLatticeAveragingGroup:2019iem} \BY{Flavour Lattice Averaging Group (FLAG)} \IN{Eur. Phys. J. C}{80}{2020}{2};
\bibitem{WilsonOPE} \BY{Wilson K.~G.} \IN{Phys. Rev.}{179}{1969}{5};
\bibitem{KadanoffOPE} \BY{Kadanoff L.~P.} \IN{Phys. Rev. Lett.}{23}{1969}{25};
\bibitem{GambinoHashimoto} \BY{Gambino P. \atque Hashimoto S.} \IN{Phys. Rev. Lett.}{125}{2020}{3};
\bibitem{HLT} \BY{Hansen M., Lupo A. \atque Tantalo N.} \IN{Phys. Rev. D}{99}{2019}{9};
\bibitem{Chebyshev} \BY{Bailas G., Hashimoto S. \atque Ishikawa T.} \IN{PTEP}{2020}{2020}{4};
\bibitem{ETM1} \BY{Frezzotti R. \atque Rossi G.~C.} \IN{Nucl. Phys. B Proc. Suppl.}{128}{2004}{193};
\bibitem{ETM2} \BY{European Twisted Mass Collaboration} \IN{Nucl. Phys. B}{887}{2014}{19};
\bibitem{JLQCD1} \BY{JLQCD Collaboration} \IN{PoS LATTICE 2016}{2017}{192};
\bibitem{JLQCD2} \BY{JLQCD Collaboration} \IN{Phys. Rev. D}{106}{2022}{5};
\bibitem{OS} \BY{Osterwalder K. \atque Seiler E.} \IN{Annals Phys.}{110}{1978}{440};

\end{thebibliography}
\end{document}